\documentclass[12pt]{article}
\usepackage{amssymb}
\usepackage{amsmath}
\usepackage{graphicx,color}

\numberwithin{equation}{section}

\begin{document}
\begin{titlepage}
%
\vspace*{10mm}
\begin{center}
\baselineskip 25pt 
{\Large\bf
Classically conformal U(1)$^\prime$ extended Standard Model and Higgs vacuum stability 
}
\end{center}
\vspace{5mm}
\begin{center}
{\large
Satsuki Oda$^{a,b,}$\footnote{satsuki.oda@oist.jp},
Nobuchika Okada$^{c,}$\footnote{okadan@ua.edu}, 
and 
Dai-suke Takahashi$^{a,b,}$\footnote{daisuke.takahashi@oist.jp}
}
\end{center}
\vspace{2mm}

\begin{center}
{\it
$^{a}$Okinawa Institute of Science and Technology Graduate School (OIST), \\ 
1919-1 Tancha, Onna-son, Kunigami-gun, Okinawa, 904-0495, Japan
\vspace{0.5cm}

$^{b}$Research Institute, Meio University, \\
1220-1 Biimata, Nago City, Okinawa, 905-8585, Japan \\
\vspace{0.5cm}

$^{c}$Department of Physics and Astronomy, University of Alabama, \\
Tuscaloosa, Alabama 35487, USA
}
\end{center}
\vspace{0.5cm}
\begin{abstract}
We consider the minimal U(1)$^\prime$ extension of the Standard Model (SM)
  with conformal invariance at the classical level, 
  where in addition to the SM particle contents, three generations of right-handed neutrinos 
  and a U(1)$^\prime$ Higgs field are introduced. 
In the presence of the three right-handed neutrinos, which are responsible for the seesaw mechanism, 
  this model is free from all the gauge and gravitational anomalies.   
The U(1)$^\prime$ gauge symmetry is radiatively broken via the Coleman-Weinberg mechanism,
  by which the U(1)$^\prime$ gauge boson ($Z^\prime$ boson) mass 
  as well as the Majorana mass for the right-handed neutrinos are generated. 
The radiative U(1)$^\prime$ symmetry breaking also induces a negative mass squared for the SM Higgs doublet
  to trigger the electroweak symmetry breaking.  
In this context, we investigate a possibility to solve the SM Higgs vacuum instability problem.  
The model includes only three free parameters (U(1)$^\prime$ charge of the SM Higgs doublet, 
  U(1)$^\prime$ gauge coupling and $Z^\prime$ boson mass),  for which we perform parameter scan, 
  and identify a parameter region resolving the SM Higgs vacuum instability.  
We also examine naturalness of the model. 
The heavy states associated with the U(1)$^\prime$ symmetry breaking contribute to the SM Higgs self-energy. 
We find an upper bound on $Z^\prime$ boson mass, $m_{Z^\prime} \lesssim 6$ TeV, 
  in order to avoid a fine-tuning severer than 10\% level. 
The $Z^\prime$ boson in this mass range can be discovered at the LHC Run-2 in the near future. 

\end{abstract}
\end{titlepage}

\section{Introduction}

The gauge hierarchy problem is one of the most important issues in the Standard Model (SM),  
  which has been motivating us to seek new physics beyond the SM for decades. 
The problem lies in the fact that quantum corrections to the self-energy 
  of the SM Higgs doublet quadratically diverges, and this divergence (cut off by some new physics scale) 
  should be canceled by a fine-tuning of the Higgs mass parameter
  when the cutoff scale is much higher than the electroweak scale, such as the Planck mass. 
Because of the chiral nature of the SM, the SM Lagrangian possesses the conformal (scale) invariance 
  at the classical level, except for the Higgs mass term.
It has been argued in \cite{Bardeen} that once the classical conformal invariance and its minimal violation 
  by quantum anomalies are imposed on the SM, it can be free from the quadratic divergences 
  and hence the gauge hierarchy problem.  
This picture fits a setup first investigated by Coleman and Weinberg ~\cite{CW}, 
  namely, a U(1) gauge theory with a massless Higgs field, 
  where the classical conformal invariance is broken by quantum corrections in the Coleman-Weinberg effective 
  potential, and the U(1) gauge symmetry is radiatively broken (Coleman-Weinberg mechanism).

Although it is tempting to apply this Coleman-Weinberg mechanism to the SM Higgs sector, 
  this cannot work with the observed values of top quark and weak boson masses, 
  since the Coleman-Weinberg potential for the SM Higgs field is found to be unbounded 
  from below \cite{Fujikawa}. 
Therefore, in order to pursue this scheme, it is necessary to extend the SM.  
Among several new physics model proposals (see, for example, \cite{CFmodels, Khoze}), 
  classically conformal B-L extended SM proposed in \cite{IOO1} is a very simple and well-motivated model. 
The B-L (baryon number minus lepton number) is a unique anomaly-free global symmetry in the SM, 
  and it can be easily gauged.  
Associated with gauging the B-L symmetry, three generation of right-handed neutrinos and 
  a B-L Higgs field are introduced to make the model free from all gauge and gravitational anomalies, 
  and to break the B-L gauge symmetry. 
Once the B-L gauge symmetry is broken, the B-L gauge field ($Z^\prime$ boson) and the right-handed (Majorana) neutrinos 
  obtain their masses.  
With the Majorana heavy neutrinos, the seesaw mechanism \cite{seesaw} is automatically implemented.   
In \cite{IOO1}, under a requirement of the classically conformal invariance, the radiative B-L symmetry breaking 
  by the Coleman-Weinberg mechanism has been investigated. 
The B-L gauge symmetry breaking also triggers the electroweak symmetry breaking 
  by generating a negative mass squared for the SM Higgs doublet. 
Naturalness of the model requires the B-L symmetry breaking scale at or below the TeV scale ~\cite{IOO1}, 
  so that the LHC Run-2 can test the model.

In this work, we first consider a generalization of the classically conformal B-L model. 
With the same new particles (three generations of right-handed neutrinos and a new Higgs field), 
  the most general gauged U(1) extension of the SM is to introduce a U(1)$^\prime$  gauge group, 
  which is defined as a linear combination of the SM $U(1)_Y$ and the U(1) B-L gauge groups. 
Thus, we investigate the classically conformal  U(1)$^\prime$  extended SM. 
U(1)$^\prime$  charge assignments for particles which are different from those in the B-L model 
  yield a drastic change in phenomenological consequences.

In this context, we consider the SM Higgs vacuum stability. 
The SM Higgs boson has been discovered at the LHC, and 
  this marks the beginning of the experimental confirmation of the SM Higgs sector. 
The observed Higgs boson mass of around 125 GeV \cite{ ATLAS2014-06, CMS2014-07} 
  (see also the recent update from a combined analysis by the ATLAS and the CMS \cite{ACcombined})
  indicates that the electroweak vacuum is unstable \cite{RGErun}, 
  since the SM Higgs quartic coupling becomes negative far below the Planck mass, 
  for the top quark pole mass $m_t=173.34 \pm 0.76$ from the combined measurements 
  by the Tevatron and the LHC experiments \cite{MtCombine}.  
This is a serious problem in our framework, since the instability of the Higgs potential 
  induces a large tree-level mass for the U(1)$^\prime$  Higgs field  through its interaction term 
  with the SM Higgs doublet field, and spoils the Coleman-Weinberg mechanism 
  for the U(1)$^\prime$  sector. 
In addition to the proposal of the classically conformal U(1)$^\prime$  extend SM, 
  the main purpose of this work is to resolve the Higgs vacuum instability in this context.

\section{Classically conformal U(1)$^{\prime}$ extended SM} \label{sec_U1prime}

\begin{table}[t]
\begin{center}
\begin{tabular}{c|ccc|rcr}
            & SU(3)$_c$ & SU(2)$_L$ & U(1)$_Y$ & \multicolumn{3}{c}{U(1)$^\prime$} \\
\hline
&&&&&&\\[-12pt]
$q_L^i$    & {\bf 3}   & {\bf 2}& $+1/6$ & $x_q$ 		& = & $\frac{1}{3}x_H + \frac{1}{6}x_\Phi$  \\[2pt] 
$u_R^i$    & {\bf 3} & {\bf 1}& $+2/3$ & $x_u$ 		& = & $\frac{4}{3}x_H + \frac{1}{6}x_\Phi$  \\[2pt] 
$d_R^i$    & {\bf 3} & {\bf 1}& $-1/3$ & $x_d$ 		& = & $-\frac{2}{3}x_H + \frac{1}{6}x_\Phi$  \\[2pt] 
\hline
&&&&&&\\[-12pt]
$\ell_L^i$    & {\bf 1} & {\bf 2}& $-1/2$ & $x_\ell$ 	& = & $- x_H - \frac{1}{2}x_\Phi$   \\[2pt] 
$\nu_R^i$   & {\bf 1} & {\bf 1}& $0$   & $x_\nu$ 	& = & $- \frac{1}{2}x_\Phi$ \\[2pt] 
$e_R^i$   & {\bf 1} & {\bf 1}& $-1$   & $x_e$ 		& = & $- 2x_H - \frac{1}{2}x_\Phi$  \\[2pt] 
\hline
&&&&&&\\[-12pt]
$H$         & {\bf 1} & {\bf 2}& $+1/2$  &  $x_H$ 	& = & $x_H$\hspace*{12.5mm}  \\ 
$\Phi$      & {\bf 1} & {\bf 1}& $0$  &  $x_\Phi$ 	& = & $x_\Phi$  \\ 
\end{tabular}
\end{center}
\caption{
Particle contents. 
In addition to the SM particle contents, the right-handed neutrino $\nu_R^i$ ($i=1,2,3$ denotes the generation index) and a complex scalar $\Phi$ are introduced. 
}
\label{Tab:particle_contents}
\end{table}

The model we will investigate is the anomaly-free U(1)$^\prime$ extension of the SM 
  with the classically conformal invariance, which is based on the gauge group 
  SU(3)$_C \times$SU(2)$_L \times$U(1)$_Y \times$U(1)$^\prime$. 
The particle contents of the model are listed in Table~\ref{Tab:particle_contents}.
The covariant derivative relevant to U(1)$_Y \times$ U(1)$^\prime$ is given by 
\begin{equation}
D_\mu \equiv \partial_\mu  
			- i ( g_1 Y + \tilde{g} Y^\prime ) B_\mu - i g^\prime Y^\prime Z^{\prime}_\mu, 
 \label{Eq:covariant_derivative}
\end{equation}
where $Y$ ($Y'$) are U(1)$_Y$ (U(1)$^\prime$ ) charge of a particle, and the gauge coupling $\tilde{g}$ 
   is introduced associated with a kinetic mixing between the two U(1) gauge bosons. 
The particle contents include three generations of right-hand neutrinos $\nu_R^i$ 
  and a U(1)$^\prime$  Higgs field $\Phi$, in addition to the SM particle contents.

For generation-independent charge assignments,  the U(1)$^\prime$ charges of the fermions 
  are defined to satisfy the gauge and gravitational anomaly-free conditions:
\begin{align}
{\rm U}(1)^\prime \times \left[ {\rm SU}(3)_C \right]^2&\ :&
			2x_q - x_u - x_d &\ =\  0, \nonumber \\
{\rm U}(1)^\prime \times \left[ {\rm SU}(2)_L \right]^2&\ :&
			3x_q + x_\ell &\ =\  0, \nonumber \\
{\rm U}(1)^\prime \times \left[ {\rm U}(1)_Y \right]^2&\ :&
			x_q - 8x_u - 2x_d + 3x_\ell - 6x_e &\ =\  0, \nonumber \\
\left[ {\rm U}(1)^\prime \right]^2 \times {\rm U}(1)_Y&\ :&
			x_q^2 - 2x_u^2 + x_d^2 - x_\ell^2 + x_e^2 &\ =\  0, \nonumber \\
\left[ {\rm U}(1)^\prime \right]^3&\ :&
			6x_q^3 - 3x_u^3 - 3x_d^3 + 2x_\ell^3 - x_\nu^3 - x_e^3 &\ =\  0, \nonumber \\
{\rm U}(1)^\prime \times \left[ {\rm grav.} \right]^2&\ :&
			6x_q - 3x_u - 3x_d + 2x_\ell - x_\nu - x_e &\ =\  0. 
\label{Eq:anomaly_free_cond}
\end{align}
In order to reproduce observed fermion masses  and flavor mixings, 
  we introduce the following Yukawa interactions: 
\begin{equation}
{\cal L}_{Yukawa} = - Y_u^{ij} \overline{q_L^i} \tilde{H} u_R^j
                                - Y_d^{ij} \overline{q_L^i} H d_R^j 
				- Y_\nu^{ij} \overline{\ell_L^i} \tilde{H} \nu_R^j - Y_e^{ij} \overline{\ell_L^i} H e_R^j
				- Y_M^i \Phi \overline{\nu_R^{ic}} \nu_R^i + {\rm h.c.},
\label{Eq:L_Yukawa}
\end{equation}
where $\tilde{H} \equiv i  \tau^2 H^*$ and the third and fifth terms in the right-handed side 
  are for the seesaw mechanism to generate neutrino masses. 
These Yukawa interaction terms impose 
\begin{eqnarray}
x_H 		&=& - x_q + x_u \ =\  x_q - x_d \ =\  - x_\ell + x_\nu \ =\  x_\ell - x_e, \nonumber \\
x_\Phi	&=& - 2x_\nu. 
\end{eqnarray} 
Solutions to these conditions are listed in Table~\ref{Tab:particle_contents}, 
  which are controlled by only two parameters, $x_H$ and $x_\Phi$.  
The two parameters reflect the fact that the U(1)$^\prime$  gauge group can be defined as a linear combination 
  of the SM U(1)$_Y$ and the U(1) B-L gauge groups. 
Since the U(1)$^\prime$  gauge coupling $g'$ is a free parameter of the model and 
  it always appears as a product $x_\Phi g^\prime$, we fix $x_\Phi=2$ without loss of generality 
  throughout this paper. 
This convention excludes the case that U(1)$^\prime$ gauge group is identical with the SM U(1)$_Y$. 
The choice of $(x_H, x_\Phi)=(0, 2)$ corresponds to the U(1)$_{B-L}$ model. 
Another example is $(x_H, x_\Phi)=(-1, 2)$, which corresponds to the SM with the so-called U(1)$_R$ symmetry. 
When we choose $(x_H, x_\Phi)=(-16/41, 2)$, the beta function of $\tilde{g}$ at the 1-loop level 
   becomes proportional to $\tilde{g}$ (see Appendix A). 
This is the orthogonal condition for the U(1)$_Y$ and U(1)$^\prime$, 
 under which $\tilde{g}$ dose not evolve once we have set $\tilde{g}=0$ at an energy scale.

Imposing the classically conformal invariance, the scalar potential is given by
\begin{equation}
V = \lambda_H \! \left( H^\dagger H \right)^2 
	+ \lambda_\Phi \! \left( \Phi^\dagger \Phi \right)^2 
	+ \lambda_{mix} \! \left( H^\dagger H \right) \! \left( \Phi^\dagger \Phi \right) , 
\label{Eq:classical_potential}
\end{equation}
where the mass terms are forbidden by the conformal invariance. 
Clearly, if  $\lambda_{mix}$ is negligibly small, 
  we can analyze the Higgs potential separately for $\Phi$ and $H$. 
This will be justified in the following sections. 
When the Majorana Yukawa couplings $Y_M^i$ is negligible compared to the U(1)$^\prime$ gauge coupling, 
 the $\Phi$ sector is identical with the original Coleman-Weinberg model \cite{CW}, 
   so that the radiative U(1)$^\prime$ symmetry breaking will be achieved. 
Once $\Phi$ develops a VEV through the Coleman-Weinberg mechanism, 
   the tree-level mass term for the SM Higgs is effectively generated through $\lambda_{mix}$ 
   in Eq.~(\ref{Eq:classical_potential}).
Taking $\lambda_{mix}$ negative, the induced mass squared for the Higgs doublet 
  is negative and, as a result, the electroweak symmetry breaking is driven in the same way as in the SM.

\section{Radiative U(1)$^\prime$ symmetry breaking}
Assuming $\lambda_{mix}$ is negligibly small, we first analyze the U(1)$^\prime$ Higgs sector. 
Without mass terms, the Coleman-Weinbeg potential \cite{CW} at the 1-loop level is found to be 
\begin{eqnarray}
  V(\phi) =  \frac{\lambda_\Phi}{4} \phi^4 
 + \frac{\beta_\Phi}{8} \phi^4 \left(  \ln \left[ \frac{\phi^2}{v_\phi^2} \right] - \frac{25}{6} \right), 
\label{Eq:CW_potential} 
\end{eqnarray}
where $\phi / \sqrt{2} = \Re[\Phi]$, and 
  we have chosen the renormalization scale to be the VEV of $\Phi$ ($\langle \phi \rangle =v_\phi$).  
Here, the coefficient of the 1-loop quantum corrections is given by 
\begin{eqnarray}
\beta_\Phi	&=& \frac{1}{16 \pi^2} 
		\left[ 20\lambda_\Phi^2 
			+ 6 x_\Phi^4 \left ( \tilde{g}^2 + g^{\prime 2} \right)^2 - 16\sum_i(Y_M^i)^4 \right] \\
	& \simeq &  \frac{1}{16 \pi^2} 
		\left[ 6 \left(x_\Phi g^\prime \right)^4 - 16\sum_i(Y_M^i)^4 \right] , 
\end{eqnarray}
 where in the last expression, we have used 
  $\lambda_\Phi^2 \ll (x_\Phi g^\prime)^4$  as usual in the Coleman-Weinberg mechanism 
   and set $\tilde{g}= 0$ at $ \langle \phi \rangle = v_\phi$,  for simplicity. 
The stationary condition $\left. dV/d\phi\right|_{\phi=v_\phi} = 0$ leads to 
\begin{eqnarray}
   \lambda_\Phi = \frac{11}{6} \beta_\Phi, 
\label{eq:stationary}
\end{eqnarray} 
 and this $\lambda_\Phi$ is nothing but a renormalized self coupling at $v_\phi$ defined as 
\begin{eqnarray}
 \lambda_\Phi = \frac{1}{3 !}\left. \frac{d^4V(\phi)}{d \phi^4} \right|_{\phi=v_\phi}. 
\end{eqnarray}  
For more detailed discussion, see \cite{Khoze}.

Associated with this radiative U(1)$^\prime$ symmetry breaking 
 (as well as the electroweak symmetry breaking),  
  the U(1)$^\prime$ gauge boson ($Z^\prime$ boson) 
  and the right-handed Majorana neutrinos acquire their masses as 
\begin{eqnarray}
  m_{Z^\prime}  = \sqrt{(x_\Phi g^\prime v_\phi)^2  +  (x_H g^\prime v_h)^2} \simeq   x_\Phi g^\prime v_\phi, 
   \;  \;  m_{N^i} = \sqrt{2} Y_M^i v_\phi, 
\end{eqnarray} 
where $v_h=246$ GeV is the SM Higgs VEV, and we have used $x_\Phi v_\phi  \gg x_H v_h$, 
  which will be verified below. 
In this paper, we assume degenerate masses for the three Majorana neutrinos,  
 $Y_M^i = y_M$ (equivalently, $m_{N^i}=m_N$) for all $i=1,2,3$, for simplicity. 
The U(1)$^\prime$ Higgs boson mass is given by 
\begin{eqnarray}
  m_\phi^2 = \left. \frac{d^2 V}{d\phi^2}\right|_{\phi=v_\phi}  
                    =\beta_\Phi v_\phi^2  \simeq 
  \frac{3}{8 \pi^2} \left( (x_\Phi g^\prime)^4 - 8 y_M^4  \right) v_\phi^2 
  \simeq  \frac{3}{8 \pi^2}  \frac{ m_{Z^\prime}^4 - 2 m_N^4}{v_\phi^2}. 
\label{Eq:mass_phi}
\end{eqnarray} 
When the Yukawa coupling is negligibly small, this reduces to the well-known relation 
  derived in the radiative symmetry breaking by the Coleman-Weinberg mechanism \cite{CW}. 
For a sizable Majorana mass, this formula indicates that 
  the potential minimum disappears for $ m_N >  m_{Z^\prime}/2^{1/4}$, 
  so that there is an upper bound on the right-handed neutrino mass 
  for the U(1)$^\prime$ symmetry to be broken radiatively.  
This is in fact the same reason as why the Coleman-Weinberg mechanism
   in the SM Higgs sector fails to break the electroweak symmetry 
   when the top Yukawa coupling is large as observed. 
In order to avoid the destabilization of the U(1)$^\prime$ Higgs potential, 
    we simply set $m_{Z^\prime}^4 \gg m_N^4$ in the following analysis. 
Note that this condition does not mean that the Majorana neutrinos must be very light, 
  even though a factor difference between $m_{Z^\prime}$ and $m_N$ is enough to satisfy the condition.

\section{Electroweak symmetry breaking}
Let us now consider the SM Higgs sector. 
In our model, the electroweak symmetry breaking is achieved in a very simple way. 
Once the U(1)$^\prime$ symmetry is radiatively broken, 
  the SM Higgs doublet mass is generated through the mixing term between $H$ and $\Phi$ 
  in the scalar potential (see Eq.~(\ref{Eq:classical_potential})), 
\begin{equation}
  V(h) = \frac{\lambda_H}{4}h^4 + \frac{\lambda_{mix}}{4} v_\phi^2 h^2,  
\end{equation}
where $H = 1/\sqrt{2}\, (0,\,h)$.
Choosing $\lambda_{mix} < 0$, the electroweak symmetry is broken in the same way as in the SM \cite{IOO1}. 
However, the crucial difference from the SM is that in our model the electroweak symmetry breaking 
   originates from the radiative breaking of the U(1)$^\prime$ gauge symmetry. 
At the tree level, the stationary condition $V^\prime |_{h=v_h} = 0$ 
   leads to the relation 
   $|\lambda_{mix}|= 2 \lambda_H (v_h/v_\phi)^2$, 
   and the Higgs boson mass $m_h$ is given by 
\begin{equation}
  m_h^2 = \left. \frac{d^2 V}{dh^2} \right|_{h=v_h} = |\lambda_{mix}|v_\phi^2 = 2 \lambda_H v_h^2. 
\label{Eq:mass_h}
\end{equation}
In the following RGE analysis, this is used as the boundary condition for $\lambda_{mix}$ 
  at the normalization scale $\mu=v_\phi$. 
Note that since $\lambda_H \sim 0.1$ and $v_\phi \gtrsim 10$ TeV by the LEP constraint \cite{LEP2A, LEP2B},  
  $|\lambda_{mix}| \lesssim 10^{-5}$, which is very small.

In our discussion about the U(1)$^\prime$ symmetry breaking, we neglected $\lambda_{mix}$ 
  by assuming it to be negligibly small. 
Here we justify this treatment. 
In the presence of $\lambda_{mix}$ and the Higgs VEV, Eq.~(\ref{eq:stationary}) is modified as 
\begin{eqnarray}
 \lambda_\Phi = \frac{11}{6} \beta_\Phi + \frac{|\lambda_{mix}|}{2} \left( \frac{v_h}{v_\phi} \right)^2 
 \simeq  
  \frac{1}{2 v_\phi^4}  \left(  \frac{11}{8 \pi^2} m_{Z^\prime}^4 + m_h^2 v_h^2    \right).   
\label{eq:consistency} 
\end{eqnarray}
Considering the current LHC bound form search for $Z^\prime$ boson resonances \cite{ATLAS_Zp, CMS_Zp},  
  $m_{Z^\prime} \gtrsim 3 $ TeV,  we find that the first term in the parenthesis 
  in the last equality is 5 orders of magnitude greater than the second term, 
  and therefore we can analyze the two Higgs sectors separately.

\section{Solving the SM Higgs vacuum instability}

\begin{figure}[t]
\begin{minipage}{0.5\linewidth}
\begin{center}
\includegraphics[width=0.95\linewidth]{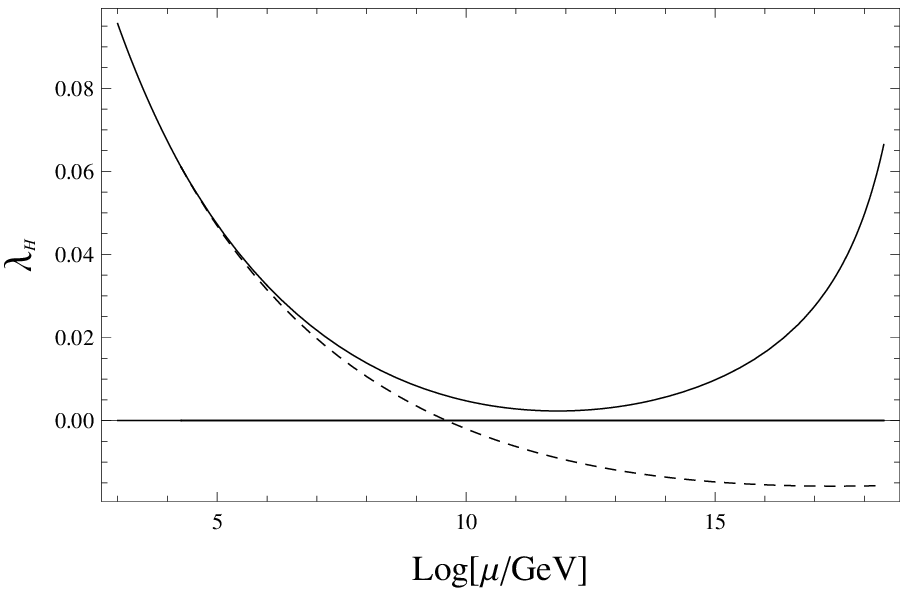}
(a)
\end{center}
\end{minipage}
\begin{minipage}{0.5\linewidth}
\begin{center}
\includegraphics[width=0.95\linewidth]{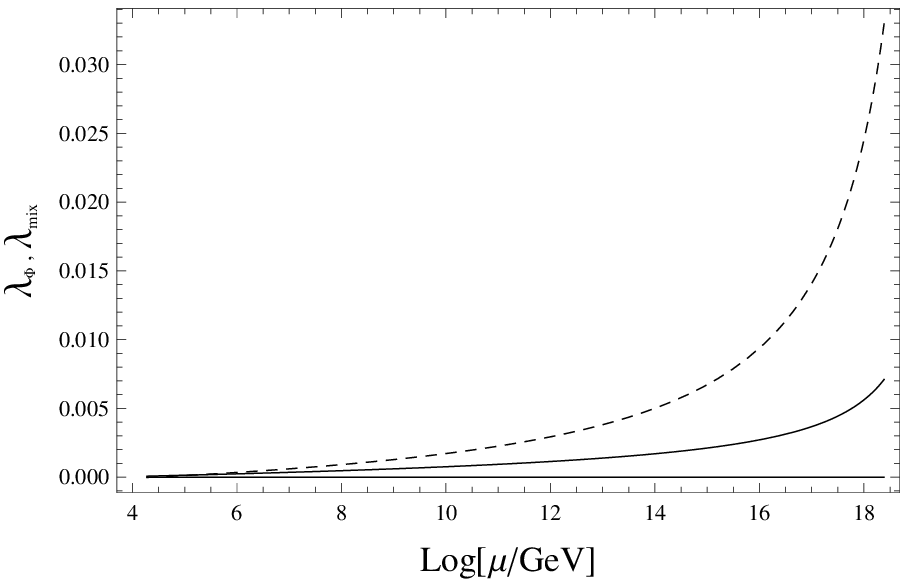}
(b)
\end{center}
\end{minipage}
\caption
{
(a) The evolutions of the Higgs quartic coupling $\lambda_H$ (solid line) 
    for the inputs $m_t=173.34$ GeV and $m_h=125.03$ GeV, 
    along with the SM case (dashed line). 
(b) The RG evolutions of $\lambda_\Phi$ (solid line) and $\lambda_{mix}$ (dashed line). 
Here, we have taken $x_H = 2$, $v_\phi = 19$ TeV and $g^\prime(v_\phi) = 0.09$. 
}
\label{Fig:Higgs_quartics}
\end{figure}

In our classically conformal U(1)$^\prime$ extended SM,  the U(1)$^\prime$ gauge symmetry 
  is radiatively broken by the Coleman-Weinberg mechanism. 
Associated with this symmetry breaking,  
  the negative Higgs mass squared is generated to break the electroweak symmetry as in the SM.  
In the SM with the observed Higgs boson mass around 125 GeV,  
  the RGE evolution of the SM Higgs quartic coupling shows that the coupling becomes negative 
  at the intermediate scale $\mu \simeq 10^{9}-10^{11}$ GeV \cite{RGErun}
  (dependently of input masses for the Higgs boson and top quark), 
  and hence the electroweak vacuum is unstable.  
This vacuum instability might not be a serious problem in the SM, since the lifetime  
  of the electroweak vacuum is much longer than the age of the Universe \cite{meta-stability}.  
However, in our model, 
  this SM Higgs vacuum instability generates a large negative mass squared 
  of $\Phi$ through the $\lambda_{mix}$ term, and hence the Coleman-Weinberg mechanism 
  in the U(1)$^\prime$ Higgs sector is spoiled.

In this section, we investigate RG evolution of the Higgs quartic coupling and a possibility  
  to solve the Higgs vacuum stability problem in our U(1)$^\prime$ extended SM.  
Without the classical conformal invariance, Ref.~\cite{Coriano} has considered the same problem 
  and identified parameter regions which can remove the Higgs vacuum instability. 
A crucial difference in our model is that because of the classical conformal invariance 
  and the symmetry breaking by the Coleman-Weinberg mechanism, 
  the initial values of $\lambda_\Phi$ and $\lambda_{mix}$ at $v_\phi$ are not free parameters. 
Therefore, it is nontrivial to resolve the Higgs vacuum instability in the present morel.   
The Higgs vacuum stability has been investigated in \cite{Khoze} for classically conformal extension of the SM 
  with an extend  gauge groups and particle contents (including a dark matter candidate).

For our RGE analysis, we employ the SM RGEs at 2-loop level \cite{RGErun}  
  from the top pole mass to the U(1)$^\prime$ Higgs VEV, and connect the RGEs 
  to those of the U(1)$^\prime$ extended SM at the 1-loop level. 
All formulas used in our analysis are listed in Appendices. 
For inputs for the Higgs boson mass and top quark pole mass, 
  we employ a central value of the CMS measurement $m_h=125.03$ GeV \cite{CMS2014-07}, 
  which is slightly smaller than the ATLAS measurement $m_h=125.36$ GeV \cite{ATLAS2014-06}, 
  while $m_t =173.34$ which is the central value of combined results of the Tevatron and the LHC 
  measurements of top quark mass \cite{MtCombine}.  
There are only 3 free parameters in our model, by which inputs at $v_\phi$ are determined: 
  $x_H$, $v_\phi$, and $g^\prime$.

In Fig.~\ref{Fig:Higgs_quartics} (a), we show the RG evolution of the SM Higgs quartic coupling 
   in our model (solid line), along with the SM case (dashed line).  
Here, we have taken $x_H= 2$, $v_\phi = 19$ TeV, and $g^\prime(v_\phi) = 0.09$ as an example.  
Recall that we have fixed $x_\Phi=2$ without loss of generality. 
The Higgs quartic coupling remains positive all the way up to the Planck mass, 
  so that the Higgs vacuum instability problem is solved. 
There are complex, synergetic effects in the coupled RGEs to resolve the Higgs vacuum instability 
  (see Appendices for RGEs). 
For example, the U(1)$_Y$ gauge coupling grows faster than the SM case 
  in the presence of the mixing gauge coupling ${\tilde g}$, 
  which makes the evolution of top Yukawa coupling decreasing faster than in the SM case. 
The evolution of the mixing gauge coupling is controlled by the U(1)$^\prime$ gauge coupling. 
Both of them are asymptotic non-free. 
The gauge couplings positively contribute to the beta function of 
  the SM Higgs quartic coupling, while the top Yukawa coupling gives a negative contribution.  
As a result, the RG evolutions of the gauge and top Yukawa couplings work 
  to change the sign of the the beta function of the SM Higgs quartic coupling 
  at $\mu \simeq 10^{12}$ GeV in Fig.~\ref{Fig:Higgs_quartics} (a). 
Fig.~\ref{Fig:Higgs_quartics} (b) shows the RG evolutions of the other Higgs quartic couplings. 
Note that the input of $\lambda_\Phi$ and $\lambda_{mix}$ are very small 
  because of the radiative gauge symmetry breaking, and the two couplings remain very small 
  even after reaching the Planck scale.   
Thus, the positive contribution of $\lambda_{mix}$ to the beta function of the SM Higgs quartic coupling is negligible. 
This is in sharp contrast to U(1) extended modes without the conformal invariance, 
   where $\lambda_{mix}$ is a free parameter and we can take its input to give a large, 
   positive contribution to the beta function, so that the Higgs vacuum instability problem 
   is relatively easier to solve.

\begin{figure}[t]
\begin{minipage}{0.5\linewidth}
\begin{center}
\includegraphics[width=0.95\linewidth]{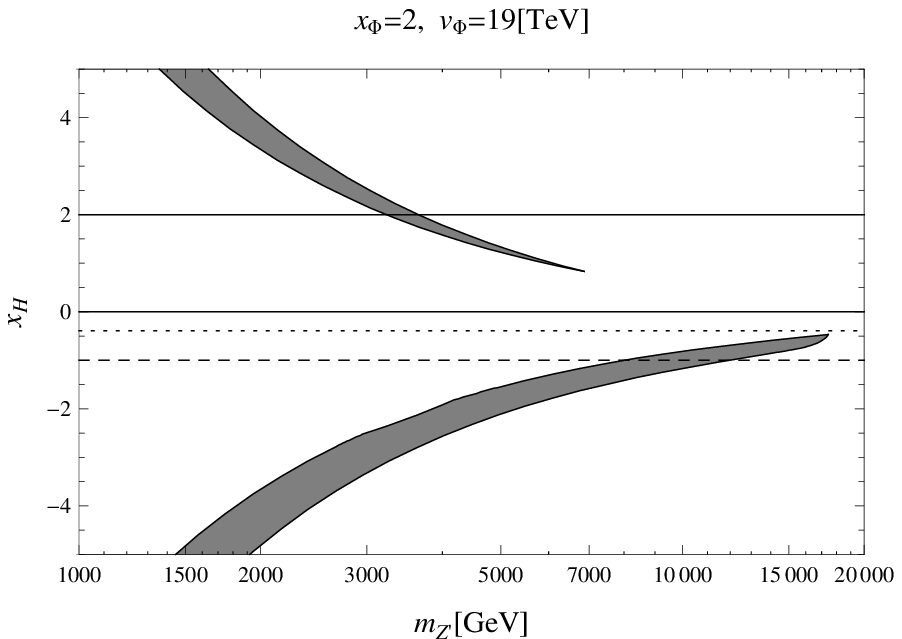}
(a)
\end{center}
\end{minipage}
\begin{minipage}{0.5\linewidth}
\begin{center}
\includegraphics[width=0.95\linewidth]{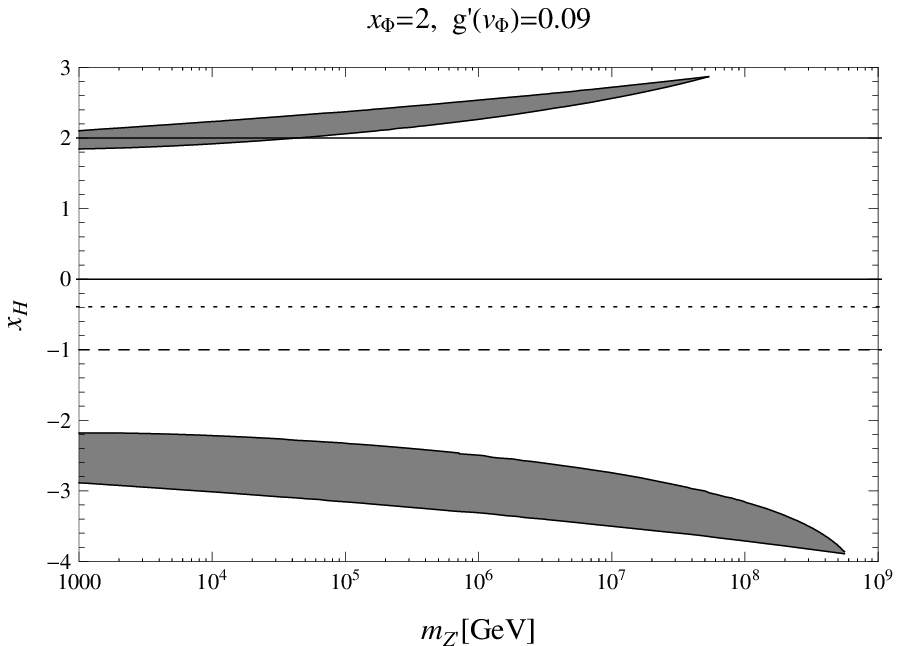}
(b)
\end{center}
\end{minipage}
\caption
{
(a) The result of parameter scan for $x_H$ and $g^\prime$ with a fixed $v_\phi=19$ TeV, 
    shown in ($m_{Z^\prime}, x_H$)-plane with 
    $m_{Z^\prime} =\sqrt{(x_\Phi g^\prime v_\phi)^2+(x_H g^\prime v_h)^2}\simeq x_\Phi g^\prime v_\phi$. 
As a reference, horizontal lines are depicted for $x_H=2$, $0$ (U(1)$_{B-L}$ case), $-16/41$ (orthogonal case),
 and $-1$ (U(1)$_R$ case). 
(b) Same as (a), but parameter scan for $x_H$ and $v_\phi$ with a fixed $g^\prime=0.09$. 
}
\label{Fig:scan1}
\end{figure}

In order to identify parameter regions to resolve the Higgs vacuum instability, 
  we also perform parameter scans for the free parameters $x_H$, $v_\phi$ and $g^\prime$. 
In this analysis, we impose several conditions on the running couplings at $v_\phi \leq \mu \leq M_P$ 
  ($M_P =2.4 \times 10^{18}$ GeV is the reduced Planck mass): 
  stability conditions of the Higgs potential  ($\lambda_H,  \lambda_\Phi > 0$), 
  and the conditions that all the running couplings remain in the perturbative regime, namely, 
  $\alpha_{g_i} \equiv g_i^2/(4\pi)<1$, $\alpha_{g^\prime} \equiv g^{\prime 2}/(4\pi)<1$,  
  $\alpha_{\tilde{g}} \equiv \tilde{g}^2/(4\pi)<1$, $\lambda_H/(4 \pi)<1$ and $\lambda_\Phi/(4 \pi) <1$.

In Fig.~\ref{Fig:scan1} we show the results of parameter scans
  for $x_H$ and $g^\prime$ with a fixed $v_\phi=19$ TeV (a),  
  and for $x_H$ and $v_\phi$ with a fixed $g^\prime=0.09$ (b), 
  in ($m_{Z^\prime}, x_H$)-plane with $m_{Z^\prime} \simeq x_\Phi g^\prime v_\phi$. 
As a reference, we also show horizontal lines corresponding to 
  $x_H=2$, $0$ (U(1)$_{B-L}$ case), $-16/41$ (orthogonal case), and $-1$ (U(1)$_R$ case). 
The resultant parameter space is very restricted. 
For example, the Higgs vacuum instability cannot be resolved in the classically conformal B-L extended SM, 
  which is also observed in \cite{Khoze}.

\begin{figure}[t]
\begin{minipage}{0.5\linewidth}
\begin{center}
\includegraphics[width=0.95\linewidth]{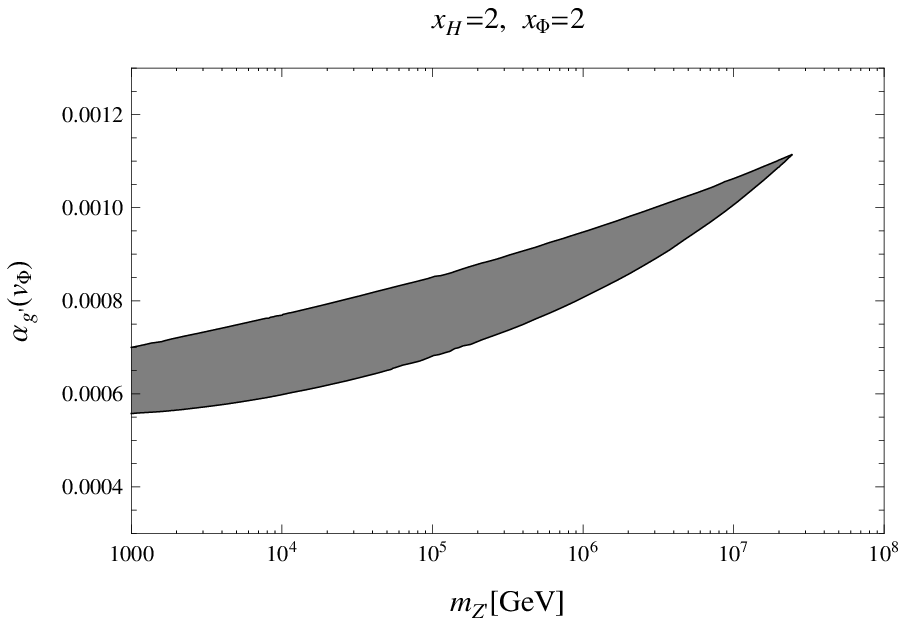}
(a)
\end{center}
\end{minipage}
\begin{minipage}{0.5\linewidth}
\begin{center}
\includegraphics[width=0.95\linewidth]{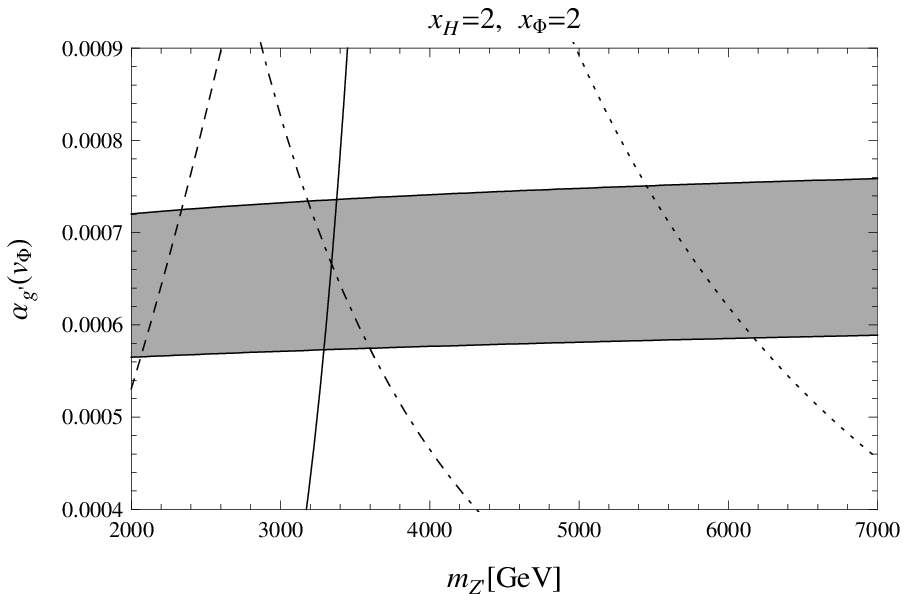}
(b)
\end{center}
\end{minipage}
\caption
{
(a) The result of parameter scan for $v_\phi$ and $g^\prime$ with a fixed $x_H=2$,   
    shown in ($m_{Z^\prime}, \alpha_{g^\prime}$)-plane.
(b) The allowed region at the TeV scale in (a) is magnified, 
    along with the LEP bound (dashed line) and the LHC bound (solid line) 
    from direct search for $Z^\prime$ boson resonance. 
    The region on the left side of the lines are excluded.    
Here, the naturalness bounds for 10\% (dotted line) and 30\% (dashed-dotted line) fine-tuning levels are also depicted. 
}
\label{Fig:scan2_xH=2}
\end{figure}

\begin{figure}[t]
\begin{minipage}{0.5\linewidth}
\begin{center}
\includegraphics[width=0.95\linewidth]{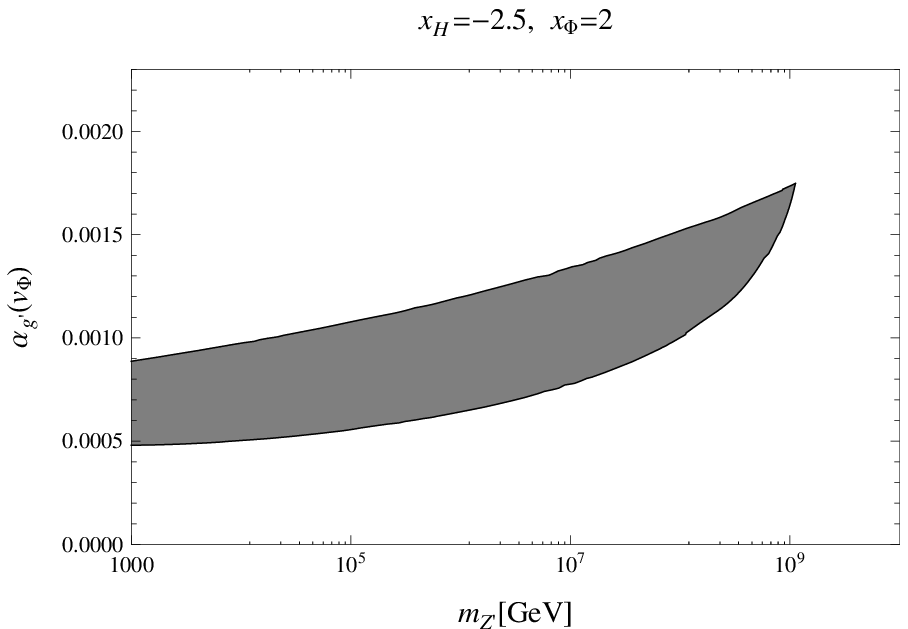}
(a)
\end{center}
\end{minipage}
\begin{minipage}{0.5\linewidth}
\begin{center}
\includegraphics[width=0.95\linewidth]{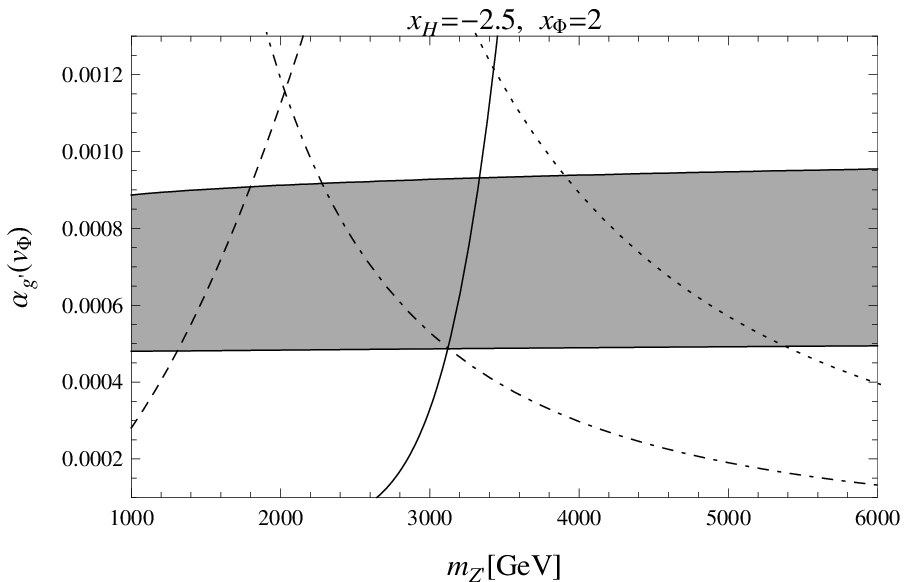}
(b)
\end{center}
\end{minipage}
\caption
{
Same as Fig.~\ref{Fig:scan2_xH=2}, but for $x_H=-2.5$. 
}
\label{Fig:scan2_xH=-2.5}
\end{figure}

The result of parameter scan for $v_\phi$ and $\alpha_{g^\prime}$ 
  with a fixed $x_H=2$ is depicted in Fig.~\ref{Fig:scan2_xH=2} (a).  
The allowed region at the TeV scale is magnified in Fig.~\ref{Fig:scan2_xH=2} (b).  
Here we also show the current collider bounds from search for $Z^\prime$ boson mediated processes 
  (details of this analysis will be presented in \cite{Okada}). 
The dashed line is obtained by interpreting the LEP results \cite{LEP2A, LEP2B} 
  for effective 4-Fermi interactions mediated by a heavy $Z^\prime$ boson, 
  while the solid line corresponds to the bound obtained by interpreting the CMS results 
  of $Z^\prime$ boson search \cite{CMS_Zp} 
  (the bound obtained from the ATLAS results \cite{ATLAS_Zp} is similar, but slightly weaker). 
The region on the left side of the lines are excluded.   
Naturalness bound, which is obtained in the next section, is also shown. 
The region, in which the classically conformal U(1)$^\prime$ model is natural,  
  will be tested by the LHC Run-2. 
Same plot but for $x_H=-2.5$ is shown in Fig.~\ref{Fig:scan2_xH=-2.5}.

\section{Naturalness bounds}

Once the classical conformal symmetry is broken and a mass scale is generated,  
   it contributes to the SM Higgs boson self-energy in general. 
Hence, if the U(1)$^\prime$ gauge symmetry breaking scale is very large, 
  we may need a fine-tuning to cancel the radiative corrections by some heavy  states 
  associated  with the U(1)$^\prime$ gauge symmetry breaking. 
  See \cite{Casas} for related discussions. 
We consider two heavy states, the right-handed neutrino and $Z^\prime$ boson, 
  whose masses are generated by the U(1)$^\prime$ gauge symmetry breaking.

Once the right-handed neutrinos obtain their Majorana masses by the U(1)$^\prime$ gauge symmetry breaking, 
  the SM Higgs self-energy is induced through the Dirac Yukawa coupling at the 1-loop level, 
  which is roughly estimated as 
\begin{equation}
 \Delta m_h^2 \sim  \frac {Y_\nu^2}{16 \pi^2} M_N^2 
  \sim \frac {m_\nu M_N^3}{16 \pi^2 v_h^2} ,
\label{Eq:contribution_YM}
\end{equation}
where we have used the seesaw formula, $m_\nu \sim Y_\nu^2 v_h^2/M_N$ \cite{seesaw}, 
  and the quadratic divergence has been dropped and the logarithmic divergence has been ignored. 
For the stability of the electroweak vacuum, we impose $\Delta m_h^2 \lesssim m_h^2$ as the naturalness. 
For example, when the light neutrino mass scale is around $m_\nu \sim 0.1$ eV, 
  we have an upper bound for the  Majorana mass as $M_N \lesssim 10^7$ GeV. 
This bound is much larger than the scale that we are interested in, $M_N \lesssim 1$ TeV.

A more important contribution to the Higgs self-energy is generated through the 1-loop diagram 
  with the $Z^\prime $ gauge boson. 
This contribution is typical in the U(1)$^\prime$ extended SM, where the SM Higgs doublet 
  has a non-zero U(1)$^\prime$ charge. 
It is in sharp contrast to the B-L extension where the SM Higgs doublet has no B-L charge, 
  and the $Z^\prime$ boson contribution arises at the 2-loop level \cite{IOO1}. 
Since the SM Higgs self-energy at the 1-loop level with the massive $Z^\prime$ boson 
  includes quadratic and logarithmic divergences, this simple calculation doesn't seem 
  consistent with our scheme for the Coleman-Weinberg mechanism. 
It may be more reasonable to calculate a series of 1-loop corrections with the $Z^\prime$ boson 
  running in the loop, with external lines of $(H^\dagger H)^n (\Phi^\dagger \Phi)^m$ ($n,m=0,1,2,\cdots$), 
  and extract a coefficient of $(H^\dagger H)$ with replacing the other lines by their VEVs.  
Since $v_h \ll v_\phi$, we expect that the dominant contributions come from 1-loop corrections 
  with external lines of $(H^\dagger H) (\Phi^\dagger \Phi)^m$.  
We may simply calculate the corrections from the Coleman-Weinberg potential in Eq.~(\ref{Eq:CW_potential}), 
\begin{eqnarray}
   V(\phi)_{\rm 1-loop} = \frac{\beta_\Phi}{8} \phi^4 \left(  \ln \left[ \frac{\phi^2}{v_\phi^2} \right] - \frac{25}{6} \right), 
\label{Eq:CW_potential_loop} 
\end{eqnarray}
   which is from 1-loop corrections with $Z^\prime$ boson running in the loop, 
   with external lines of $(\Phi^\dagger \Phi)^m$. 
Here we ignore $Y_M^i$ in $\beta_\Phi$.\footnote{
   We may calculate the contributions from the right-handed neutrinos in the same way 
   by using $Y_M$s, although we have concluded from the rough estimate of Eq.~(\ref{Eq:contribution_YM})  
   that the contributions are not significant. }  
With the 1-loop corrections, we replace one combination of external line $(\Phi^\dagger \Phi)$ 
  and a corresponding vertex $ (x_\Phi g^\prime)^2 $ 
   to $(H^\dagger H)$ with its vertex $ (x_H g^\prime)^2$. 
Thus, we evaluate the SM Higgs self-energy by 
\begin{eqnarray}
 \Delta m_h^2 = \left. \frac{d V_{\rm 1-loop}}{d (\phi^2)} \right|_{\phi^2=v_\phi^2} 
   \times \frac{x_H^2}{x_\Phi^2}  \times 2 
  = - \frac{11}{4 \pi} x_H^2 \; \alpha_{g^\prime} \; m_{Z^\prime}^2 . 
\label{Eq:Z'corrections}
\end{eqnarray}

If $\Delta m_h^2$ is much larger than the electroweak scale,  
  we need a fine-tuning of the tree-level Higgs mass ($|\lambda_{mix}| v_\phi^2/2$) 
  to reproduce the correct value for the SM Higgs VEV, $v_h$. 
We simply evaluate a fine-tuning level as 
\begin{eqnarray}
  \delta  = \frac{m_h^2}{2 |\Delta m_h^2|}. 
\end{eqnarray}
Here, $\delta =0.1$, for example, indicates that we need to fine-tune the tree-level Higgs mass squared 
   at the accuracy of 10\% level.  
Some of fine-tuning levels are shown in Figs.~\ref{Fig:scan2_xH=2} and \ref{Fig:scan2_xH=-2.5}, 
   along  with the results of parameter scans.

\section{Conclusions}
The classical conformal symmetry with its violation through quantum anomalies 
  could be a solution to the gauge hierarchy problem in the SM. 
Because of the absence of the mass term in the Higgs potential in this system, 
  the gauge symmetry breaking should be radiatively induced by the Coleman-Weinberg mechanism.   
Unfortunately, we cannot simply apply this mechanism to the SM, since the large top Yukawa coupling 
  destabilizes the effective Higgs potential.  
We have extended the SM by introducing an anomaly-free U(1)$^\prime$ symmetry, 
  along with three right-handed neutrinos and a U(1)$^\prime$ Higgs field. 
The U(1)$^\prime$ symmetry is radiatively broken by the Coleman-Weinberg mechanism, 
  by which the $Z^\prime$ boson as well as the right-handed neutrinos acquire their masses. 
Through a mixing terms between the U(1)$ ^\prime$ Higgs and the SM Higgs doublet fields, 
  a negative mass squared for the SM Higgs doublet is generated and, as a result,   
  the electroweak symmetry breaking is triggered.  
Therefore, all mass generations occur through the dimensional transmutation.

In the context of the classically conformal U(1)$^\prime$ model, 
  we have investigated a possibility to resolve the SM Higgs vacuum instability. 
Since the gauge symmetry is broken by the Coleman-Weinberg mechanism, 
  all quartic couplings in the Higgs potential except the SM Higgs one are very small, 
  and hence their positive contributions to U(1)$^\prime$ model,
  are not effective in resolving the SM Higgs vacuum instability.  
On the other hand, in the U(1)$^\prime$ model, the SM Higgs doublet has a non-zero U(1)$^\prime$ charge, 
  and this gauge interaction positively contributes to the beta function. 
In addition, the U(1)$^\prime$ gauge interaction negatively contributes to the beta function of  
  the top Yukawa coupling, so that the running top Yukawa coupling is decreasing faster 
  than in the SM case, and its negative contribution to the beta function 
  of the SM Higgs quartic coupling becomes milder.  
For three free parameters of the model, we have performed parameter scan, 
  and found a parameter region to solve the SM Higgs vacuum instability problem.

We have also considered naturalness of our model. 
After the U(1)$^\prime$ gauge symmetry breaking, the heavy states, $Z^\prime$ boson 
  and the right-handed neutrinos, contribute to the SM Higgs self-energy. 
Therefore, the self-energy exceeds the electroweak scale, if the states are too heavy. 
Since the SM Higgs doublet has non-zero U(1)$^\prime$ charge, 
  this self-energy corrections from $Z^\prime$ boson occur at the one loop level. 
This is in sharp contrast with the classically conformal B-L model~\cite{IOO1},  
  where the Higgs doublet has no B-L charge, and the self-energy corrections 
  from $Z^\prime$ boson occur at the two loop level.  
The naturalness constraint leads to the upper bound on the $Z^\prime$ boson mass 
  as $m_{Z^\prime} \lesssim 6$ TeV. 
The $Z^\prime$ boson with this mass range will be tested at the LHC Run-2 in the near future.

\section*{Acknowledgements}
We would like to thank H.~Sugawara, S.~Hikami, H.~Shimada, H.~Ueda, N.~Haba, and S.~Okada 
  for discussions and useful comments.  
S.O. and D.T. acknowledge Department of Physics and Astronomy, University of Alabama, 
  for their hospitality during their visit. 
The work of N.O. is supported in part by the United States Department of Energy.

\appendix
\section{The U(1)$^\prime$ RGEs at one-loop level} 
\label{Sec:U(1)'_RGEs}
In this appendix we present the one-loop RGEs for the U(1)$^\prime$ extension 
   of the SM, which are used in our analysis.  
The definitions of the covariant derivative, the Yukawa interactions and the scalar potential are given 
   by Eqs.~(\ref{Eq:covariant_derivative}), (\ref{Eq:L_Yukawa}) and (\ref{Eq:classical_potential}), respectively.
We only include the top quark Yukawa coupling $y_t$ and the right-handed neutrino Majorana Yukawa coupling $Y_M^i$, 
   since the other Yukawa couplings are negligibly small. 
The U(1)$^\prime$ charges $x_i$ are defined in Table~\ref{Tab:particle_contents}. 
The RGEs for the gauge couplings at the one-loop level are given by  
\begin{eqnarray}
\mu \frac{d g_3}{d\mu} &=& \frac{g_3^3}{(4\pi)^2} \Big[ -7 \Big], 
\nonumber \\
\mu \frac{d g_2}{d\mu} &=& \frac{g_2^3}{(4\pi)^2} \left[ -\frac{19}{6} \right],  
\nonumber \\  
\mu \frac{d g_1}{d\mu} &=& \frac{g_1}{(4\pi)^2} 
	\left[ 12\left(\frac{1}{6}g_1+x_q\tilde{g} \right)^{\!\!2} + 6\left(\frac{2}{3}g_1+x_u\tilde{g} \right)^{\!\!2} 
		+ 6\left(-\frac{1}{3}g_1+x_d\tilde{g} \right)^{\!\!2} 
	\right. \nonumber \\
	&+& \left.
	 4\left(-\frac{1}{2}g_1+x_\ell\tilde{g} \right)^{\!\!2}
		+ 2\left(x_\nu\tilde{g} \right)^2 + 2\left(-g_1+x_e\tilde{g} \right)^2
	     + \frac{2}{3}\left(\frac{1}{2}g_1+x_H\tilde{g} \right)^{\!\!2} 
		+ \frac{1}{3}\left(x_\Phi\tilde{g} \right)^2 \right], 
 \nonumber \\  
\mu \frac{d g^\prime}{d\mu} &=& \frac{g^{\prime 3}}{(4\pi)^2} 
	\left[ 12x_q^2 + 6x_u^2 + 6x_d^2 + 4x_\ell^2+ 2x_\nu^2 + 2x_e^2
		 + \frac{2}{3}x_H^2 + \frac{1}{3}x_\Phi^2 \right], 
    \nonumber \\  
\mu \frac{d \tilde{g}}{d\mu} &=& \frac{1}{(4\pi)^2} 	\left[ \tilde{g} \left\{ 
		12\left(\frac{1}{6}g_1+x_q\tilde{g} \right)^{\!\!2} + 6\left(\frac{2}{3}g_1+x_u\tilde{g} \right)^{\!\!2} 
		+ 6\left(-\frac{1}{3}g_1+x_d\tilde{g} \right)^{\!\!2} 
	\right. \right. \nonumber \\
	&+&  4\left(-\frac{1}{2}g_1+x_\ell\tilde{g} \right)^{\!\!2}
		+ 2\big(x_\nu\tilde{g} \big)^2 + 2\big( - g_1 + x_e\tilde{g} \big)^2
	\left.
		 + \frac{2}{3}\left(\frac{1}{2}g_1+x_H\tilde{g} \right)^{\!\!2} 
		+ \frac{1}{3}\left(x_\Phi\tilde{g} \right)^2 \right\} \nonumber \\
	&+& 2g^{\prime 2} \!
	\left\{ 12x_q \! \left(\frac{1}{6}g_1+x_q\tilde{g} \right) + 6x_u \!\left(\frac{2}{3}g_1+x_u\tilde{g} \right) 
		+ 6x_d \!\left(-\frac{1}{3}g_1+x_d\tilde{g} \right) 
	\right. \nonumber \\
	&+& 4x_\ell\left(-\frac{1}{2}g_1+x_\ell\tilde{g} \right)
		+ 2x_\nu \big(x_\nu \tilde{g} \big) + 2x_e\big(-g_1+x_e\tilde{g} \big)
	\nonumber \\
	&+&\left. \left.
		 \frac{2}{3}x_H \left(\frac{1}{2}g_1+x_H\tilde{g} \right) 
		+ \frac{1}{3}x_\Phi \left(x_\Phi\tilde{g} \right) \right\} \right]. 
\label{Eq:RGE_gauge}  
\end{eqnarray}
For the RGEs for the Yukawa couplings at the one-loop level we have 
\begin{eqnarray}
\mu \frac{d y_t}{d\mu} &=& \frac{y_t}{(4\pi)^2} 
	\left[ \frac{9}{2}y_t^2 - 8g_3^2 - \frac{9}{4}g_2^2 
		- 6\left( \frac{1}{6}g_1+x_q\tilde{g} \right) \left( \frac{2}{3}g_1+x_u\tilde{g} \right) 
	\right. 
\nonumber \\ 
  &-& 3 \left(\frac{1}{2}g_1+x_H\tilde{g} \right)^{2} 
	\Bigg.- 6 \left( x_q g^\prime \right)  \left( x_u g^\prime \right) 
		- 3 \left(x_H g^\prime \right)^2 \Bigg],   \nonumber \\
\mu \frac{d Y_M^i}{d\mu} &=& \frac{Y_M^i}{(4\pi)^2} 
	\Bigg[ 4(Y_M^i)^2 + 2\sum_j(Y_M^j)^2 
			+ \left( 6x_\nu^2 - 3x_\Phi^2 \right) \! \left( \tilde{g}^2 + g^{\prime 2} \right) \Bigg]. 
\label{Eq:RGE_Yukawa}  
\end{eqnarray} 
Finally, the RGEs for the scalar quartic couplings are given by 
\begin{eqnarray}
\mu \frac{d \lambda_H}{d\mu} &=& \frac{1}{(4\pi)^2} 
	\left[ \lambda_H \left\{ 24\lambda_H + 12y_t^2 - 9g_2^2 - 12\left(\frac{1}{2}g_1+x_H\tilde{g} \right)^{2} 
						- 12\big( x_H g^\prime \big)^2 \right\}
	\right. \nonumber \\
	&+& \left.
	    \lambda_{mix}^2 - 6y_t^4 
		+ \frac{9}{8}g_2^4 + 6\left(\frac{1}{2}g_1+x_H\tilde{g} \right)^{\!\!4} + 6\big( x_H g^\prime \big)^4
	\right. \nonumber \\
	&+&\left.
		 3g_2^2\left( \frac{1}{2}g_1 + x_H\tilde{g} \right)^{2} 
		+ 3g_2^2\big ( x_H g^\prime \big)^2 
		+ 12 \left( \frac{1}{2}g_1 + x_H\tilde{g} \right)^{\!\!2} \!\! \big( x_H g^\prime \big)^2 
	 \right],  
	 \nonumber \\
\mu \frac{d \lambda_\Phi}{d\mu} &=& \frac{1}{(4\pi)^2} 
	\Bigg[ \lambda_\Phi \Big\{ 20\lambda_\Phi + 8\sum_i(Y_M^i)^2 
							- 12\big( x_\Phi \tilde{g} \big)^2 - 12\big( x_\Phi g^\prime \big)^2 \Big\} 
	\Bigg. \nonumber \\
	&+&\Bigg.
		 2\lambda_{mix}^2 - 16\sum_i(Y_M^i)^4 
		+ 6 \Big\{ \big( x_\Phi \tilde{g} \big)^2 + \big( x_\Phi g^\prime \big)^2 \Big\}^2 
	\Bigg], 
	 \nonumber \\  
\mu \frac{d \lambda_{mix}}{d\mu} &=& \frac{1}{(4\pi)^2} 
	\left[ \lambda_{mix} \Bigg\{ 12\lambda_H + 8\lambda_\Phi + 4\lambda_{mix} 
							+ 6y_t^2 + 4\sum_i(Y_M^i)^2 \Bigg. 
	\right. \nonumber \\
	&-& \left.
		 \frac{9}{2}g_2^2 - 6\left( \frac{1}{2}g_1+x_H\tilde{g} \right)^{\!\!2} - 6\big(x_\Phi\tilde{g} \big)^2 
		- 6\big( x_H g^\prime \big)^2 - 6\big( x_\Phi g^\prime \big)^2 
	 \right\} \nonumber \\
	&+& \left.
		12 \left\{ \left( \frac{1}{2}g_1 + x_H\tilde{g} \right) \!\! \big( x_\Phi \tilde{g} \big) 
						+ \big( x_H g^\prime \big) \! \big( x_\Phi g^\prime \big) \right\}^{2} 
	\right].
\label{Eq:RGE_lambda}
\end{eqnarray}

\section{The SM RGEs at two-loop level}
\label{Sec:SM_RGEs}
The RGEs for coupling constants of the SM up to two-loop level \cite{RGErun} are give by 
\begin{eqnarray}
\mu \frac{d g_3}{d\mu} &=& \frac{g_3^3}{(4\pi)^2} \Big[ -7 \Big]
	+ \frac{g_3^3}{(4\pi)^4} \left[ - 26g_3^2   + \frac{9}{2}g_2^2 + \frac{11}{6}g_1^2 - 2y_t^2 \right],
 \nonumber \\	
\mu \frac{d g_2}{d\mu} &=& \frac{g_2^3}{(4\pi)^2} \left[ -\frac{19}{6} \right]
	+ \frac{g_2^3}{(4\pi)^4} \left[12g_3^2 + \frac{35}{6}g_2^2 + \frac{3}{2}g_1^2 - \frac{3}{2}y_t^2 \right],
 \nonumber \\
\mu \frac{d g_1}{d\mu} &=& \frac{g_1^3}{(4\pi)^2} \left[ \frac{41}{6} \right] 
	+ \frac{g_1^3}{(4\pi)^4} 
		\left[ \frac{44}{3}g_3^2 + \frac{9}{2}g_2^2 + \frac{199}{18}g_1^2 - \frac{17}{6}y_t^2 \right],
  \nonumber\\
\mu \frac{d y_t}{d\mu} &=& \frac{y_t}{(4\pi)^2} 
		\left[ \frac{9}{2}y_t^2 - 8g_3^2 - \frac{9}{4}g_2^2 - \frac{17}{12}g_1^2 \right] \nonumber\\
	&+& \frac{y_t}{(4\pi)^4} \left[ 
			y_t^2 \! \left( - 12y_t^2 - 12\lambda_H 
						+ 36g_3^2 + \frac{225}{16}g_2^2 + \frac{131}{16}g_1^2 \right) 
		\right. \nonumber\\
	&+& \left.
			6\lambda_H^2 - 108 g_3^4 - \frac{23}{4}g_2^4 + \frac{1187}{216}g_1^4 
			+ 9g_3^2 g_2^2 + \frac{19}{9}g_3^2 g_1^2- \frac{3}{4}g_2^2 g_1^2 
	 	\right], 
	\nonumber\\
\mu \frac{d \lambda_H}{d\mu} &=& \frac{1}{(4\pi)^2} 
		\left[ \lambda_H \Big( 24\lambda_H + 12y_t^2 - 9g_2^2 - 3g_1^2 \Big)
			- 6y_t^4 + \frac{9}{8}g_2^4 + \frac{3}{8}g_1^4 + \frac{3}{4}g_2^2 g_1^2 
		\right] \nonumber\\
	&+& \frac{1}{(4\pi)^4} \left[ 
			\lambda_H^2 \Big( - 312\lambda_H - 144y_t^2 + 108g_2^2 + 36g_1^2 \Big)  
			\right. \nonumber\\
	&+& \lambda_H y_t^2 \! \left( - 3y_t^2 + 80g_3^2 + \frac{45}{2}g_2^2 + \frac{85}{6}g_1^2 \right) 
			+ \lambda_H \! \left( - \frac{73}{8}g_2^4 + \frac{629}{24}g_1^4 + \frac{39}{4}g_1^2 g_2^2 \right) 
			\nonumber\\
	&+& y_t^4 \! \left( 30y_t^2 - 32g_3^2 - \frac{8}{3}g_1^2 \right) 
			+ y_t^2 \! \left( - \frac{9}{4}g_2^4 - \frac{19}{4}g_1^4 + \frac{21}{2}g_2^2 g_1^2 \right)
			\nonumber\\
	&+& \left.
		 \frac{305}{16}g_2^6 - \frac{379}{48}g_1^6 
			- \frac{289}{48}g_2^4 g_1^2 - \frac{559}{48}g_2^2 g_1^4 
		\right] .
\label{Eq:SM_RGE_lambda_H}
\end{eqnarray}
In our analysis, we numerically solve these SM RGEs with the following boundary conditions at $\mu=m_t$ 
  \cite{RGErun} \footnote{
We employed the boundary conditions in arXiv:1307.3536v4.
} 
\begin{eqnarray}
g_3(m_t) &=& 1.1666 + 0.00314 \left( \frac{\alpha_3(m_Z) - 0.1184}{0.0007} \right) 
			-  0.00046 \left( \frac{m_t}{\rm GeV} - 173.34 \right), \nonumber \\
g_2(m_t) &=& 0.64779 + 0.00004 \left( \frac{m_t}{\rm GeV} - 173.34 \right) 
			+ 0.00011 \left( \frac{m_W - 80.384{\rm GeV}}{0.014{\rm GeV}} \right), \nonumber \\
g_1(m_t) &=& 0.35830 + 0.00011 \left( \frac{m_t}{\rm GeV} - 173.34 \right) 
			- 0.00020\left(  \frac{m_W - 80.384{\rm GeV}}{0.014{\rm GeV}} \right),\nonumber \\
y_t(m_t) &=& 0.93690 + 0.00556 \left( \frac{m_t}{\rm GeV} - 173.34 \right) 
			- 0.00042 \left( \frac{\alpha_3(m_Z) - 0.1184}{0.0007} \right), \nonumber \\
\lambda_H(m_t) &=& 0.12604 + 0.00206 \left( \frac{m_h}{\rm GeV} - 125.15 \right) 
			- 0.00004 \left( \frac{m_t}{\rm GeV} - 173.34\right), 
\label{Eq:SM_BC_lambda_H}
\end{eqnarray}
using the inputs, $\alpha_3(m_Z) = 0.1184$, $m_t=173.34$ GeV, $m_h=125.03$ GeV, 
and $m_W=80.384$ GeV. 



\begin{thebibliography}{99}
\bibitem{CMS2014-07} 
CMS Collaboration [CMS Collaboration],
 ``Precise determination of the mass of the Higgs boson
  and studies of the compatibility of its couplings with the standard model,''
  CMS-PAS-HIG-14-009.
\end{thebibliography}
\end{document}